\documentclass[amsmath,floatfix,twocolumn,superscriptaddress]{revtex4-1}
\usepackage{amssymb,amsmath}
\usepackage{graphicx,array}
\usepackage{dcolumn}
\usepackage{psfrag}
\usepackage{bm}
\usepackage{color}

\begin{document}
\title{Influence of lattice dynamics on lithium-ion conductivity: A first-principles study}

\author{Arun K. Sagotra}
\affiliation{School of Materials Science and Engineering, UNSW Australia, Sydney NSW 2052, Australia} 

\author{Dewei Chu}
\affiliation{School of Materials Science and Engineering, UNSW Australia, Sydney NSW 2052, Australia} 

\author{Claudio Cazorla}
\thanks{Corresponding Author}
\affiliation{School of Materials Science and Engineering, UNSW Australia, Sydney NSW 2052, Australia}

\begin{abstract}
In the context of novel solid electrolytes for solid-state batteries, first-principles calculations 
are becoming increasingly more popular due to their ability to reproduce and predict accurately the 
energy, structural, and dynamical properties of fast-ion conductors. In order to accelerate 
the discovery of new superionic conductors is convenient to establish meaningful relations between ionic 
transport and simple materials descriptors. Recently, several experimental studies on lithium fast-ion 
conductors have suggested a correlation between lattice softness and enhanced ionic conductivity due 
to a concomitant decrease in the activation energy for ion migration, $E_{a}$. In this article, we employ 
extensive \emph{ab initio} molecular dynamics simulations based on density functional theory to substantiate 
the links between ionic transport and lattice dynamics in a number of structurally and chemically distinct 
lithium superionic conductors. Our first-principles results show no evidence for a direct and general 
correlation between $E_{a}$, or the hopping attempt frequency, and lattice softness. However, we find
that, in agreement with recent observations, the pre-exponential factor of lithium diffusivity, $D_{0}$, follows 
the Meyer-Neldel rule $\propto \exp{\left(E_{a}/\langle \omega \rangle\right)}$, where $\langle \omega 
\rangle$ represents an average phonon frequency. Hence, lattice softness can be identified with enhanced 
lithium diffusivity but only within families of superionic materials presenting very similar migration 
activation energies, due to larger $D_{0}$. On the technical side, we show that neglection of temperature 
effects in first-principles estimation of $E_{a}$  may lead to huge inaccuracies of $\sim 10$\%. The 
limitations of zero-temperature harmonic approaches in modeling of lithium-ion conductors are also 
illustrated. 
\end{abstract}
\maketitle

\section{Introduction}
\label{sec:intro}
Fast-ion, or superionic, conductors (FIC) exhibit large ionic conductivies ($\sim 1$~mS~cm$^{-1}$) 
in the crystal phase \cite{hull04}. Examples of archetypal FIC are CaF$_{2}$, AgI, and La$_{0.5}$Li$_{0.5}$TiO$_{3}$ 
\cite{cazorla14,keen96,inaguma93}. In addition to their fundamental interests, FIC are of tremendous importance 
in technological applications such as solid-state batteries \cite{bachman16}, solid oxide fuel cells \cite{ormerod03}, 
solid-state cooling \cite{cazorla16,cazorla17a,cazorla17b,cazorla18}, and catalysis and sensors \cite{montini16,goodenough97}. 
In the context of electrochemical energy storage, lithium FIC are crucial in their role as solid electrolytes, 
which enable the back-and-forth passage of lithium ions between electrodes. Lithium FIC, however, are complex 
materials in which ionic conductivity depends strongly on their chemical composition and atomic structure, and thus far 
only a reduced number of fast-ion conductors have been identified as suitable for applications \cite{bachman16}. 
To design novel lithium FIC with enhanced ionic conductivities then is desirable to establish meaningful relationships 
between lithium diffusivity and simple materials descriptors \cite{wang15,xiao15}.

Recent studies have explored the correlations between lattice dynamics and ionic transport in 
lithium and other families of FIC \cite{goel14,krauskopf17,cazorla18b,fang17,kraft17,muy18}. 
In particular, Kraft {\emph et al.} have investigated the superionic argyrodites Li$_{6}$PS$_{5}$X 
(X = Cl, Br, I) \cite{kraft17} and Muy {\emph et al.} the LISICON series originated by Li$_{3}$PO$_{4}$ 
\cite{muy18} by using electrochemical impedance spectroscopy and neutron diffraction measurements. 
The authors of both studies have concluded that lattice softness is correlated with low activation 
energies for ion migration, $E_{a}$. The intuitive explanation for such an effect is that low-frequency 
lattice excitations involve large atomic displacements around the equilibrium positions, which may enhance 
the probability of lithium ions to hop towards adjacent sites \cite{bachman16,muy18,wakamura97}. 

On a general scale, it would be very interesting to ascertain whether the same interplay between
$E_{a}$ and lattice dynamics applies also to other families of lithium FIC that present markedly 
different compositions and structural traits (e.g., cubic antiperovskites and hexagonal nitride 
compounds --we recall that Li$_{6}$PS$_{5}$X and Li$_{3}$PO$_{4}$-based compounds mostly exhibit 
orthorhombic crystal symmetry--). Meanwhile, the intuitive explanation that has been proposed
to understand the influence of phonons on lithium-ion conductivity might be too simplistic. 
For instance, lithium ions are lightweight hence the low-frequency lattice excitations in FIC, 
which mostly are related to the mechanical stiffness of the material, generally will be dominated 
by heavier atomic species; intuitively then it could be argued that wide anion lattice vibrations 
would reduce the excursions of lithium ions (e.g., by distorting the usual low-dimensional 
ion conducting channels \cite{he17}) and thus obstruct, rather than enhance, their diffusivity. 
Moreover, large lithium displacements around the equilibrium positions involve also low vibrational 
frequencies, which suggests a reduction in the corresponding hopping attempt frequency; this effect 
would have an opposite impact on the ionic conductivity than an eventual decrease in $E_{a}$, 
and \emph{a priori} it is not clear which of the two mechanisms would be dominant \cite{kraft17}. 
For an improved design of lithium FIC, a more general and quantitative understanding of how lattice 
dynamics and ionic transport are related is crucially needed. 

\begin{figure}
\centerline
        {\includegraphics[width=1.00\linewidth]{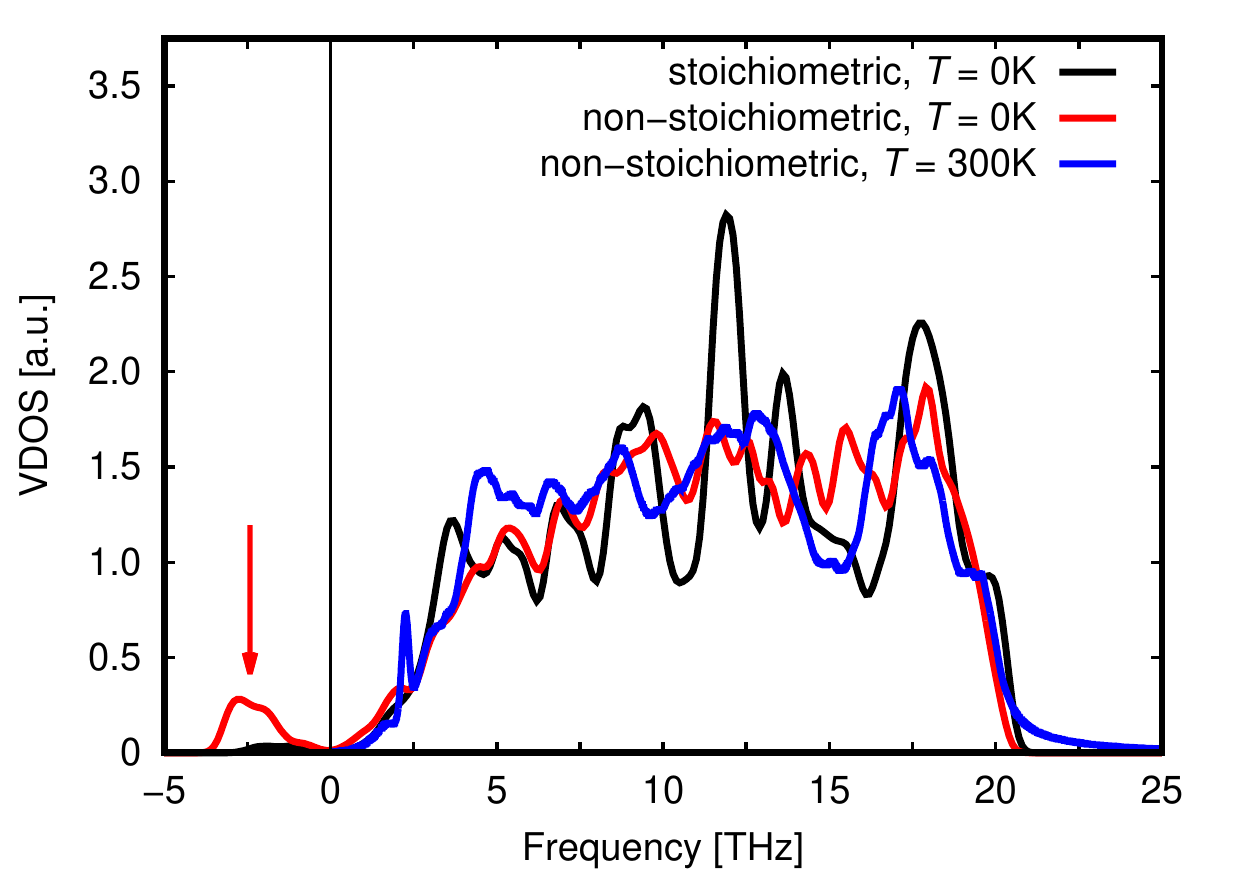}}
	\caption{Vibrational density of states of Li$_{3}$N calculated at different compositions and temperatures.
	(\emph{Black})~Stoichiometric system at zero temperature is vibrationally unstable since exhibits 
	few imaginary phonon frequencies; results are obtained with the harmonic approximation. 
	(\emph{Red})~Non-stoichiometric system at zero temperature is vibrationally unstable since exhibits 
	many imaginary phonon frequencies, most of which are associated to Li-dominated lattice eigenmodes 
        (indicated by the red arrow); 
	results are obtained with the harmonic approximation. (\emph{Blue})~Non-stoichiometric system at $T = 300$~K 
	is vibrationally stable due to the lack of imaginary phonon frequencies; results are obtained with 
	AIMD simulations, which fully take into consideration anharmonicity and temperature effects.}
\label{fig1}
\end{figure}

First-principles simulations may help at improving our comprehension of FIC via accurate estimation 
of ion-migration energy barriers, relevant thermodynamic properties, and preferred diffusion paths 
\cite{mo12,zhang13,cazorla18c}. Nevertheless, due to the intense computational expense associated 
to first-principles calculations, most quantum studies on FIC generally neglect temperature effects. 
Unfortunately, this simplification may lead to important bias and erroneous interpretations. For instance, 
zero-temperature calculations of ion-migration energy barriers customarily are performed with the 
nudged-elastic band (NEB) method \cite{henkelman00}, in which the initial and final geometries of the 
vacancy or interstitial ions need to be guessed in the form of high-symmetry metastable states; the 
limitations of this method for determining preferred ion diffusion paths are well documented for some 
prototype FIC like metal halogens (e.g., CaF$_{2}$ \cite{cazorla18c,cazorla17} and PbF$_{2}$ \cite{hutchings91,castiglione01}), 
copper chalcogenides (e.g., Cu$_{2}$S \cite{wang12}) and lithium-based oxides (e.g., LiFePO$_{4}$ \cite{yang11}).  
Likewise, phonon calculations customarily are performed at zero temperature by using the harmonic 
approximation \cite{cazorla17c,cazorla13,cazorla17d} and considering perfectly stoichiometric systems 
(i.e., full Li occupancy); such simplifications may result in a misrepresentation of real lithium FIC, 
in which sizable ionic conductivities normally appear at high temperatures and in non-stoichiometric compounds
(i.e., partial Li occupancy) \cite{bachman16,muy18}. Actually, superionic phases in lithium FIC tend to be 
highly anharmonic and become entropically stabilized at $T \neq 0$ conditions (that is, imaginary phonon 
frequencies usually appear in the corresponding zero-temperature phonon spectra), as we show in Fig.~\ref{fig1} 
(see also Supplementary Fig.1 and Refs.\cite{cazorla14,cazorla18b,muy18,chen15}). It thus seems apparent that 
considering temperature effects in first-principles simulations of FIC is actually necessary for 
better understanding them. 

\emph{Ab initio} molecular dynamics (AIMD) simulations naturally account for temperature and anharmonic
effects in materials, and thus are a powerful tool for analyzing in detail and with reliability ionic 
diffusion processes in FIC \cite{cazorla14,klerk18,singh18,richards16,kahle18}. Estimation of key quantities 
like jump rates, hopping attempt frequencies, correlation factors, and $T$--dependent phonon frequencies, 
which are not accessible with zero-temperature methods, can be obtained directly from AIMD simulations. 
The superior performance of AIMD methods certainly comes with a significant increase in computational 
expense; however, due to the current steady growth in computational power, improved design of algorithms, 
and the fact that lithium FIC typically can be described with a relatively small number of valence electrons 
by using pseudopotential approaches (in contrast, for instance, to oxide perovskites containing transition 
metals), reliable AIMD simulation of fast-ion materials is currently within reach (see works 
\cite{cazorla14,klerk18,singh18,richards16,kahle18} and Supplementary Methods).  

In this article, we present a thorough study on the lattice dynamics and ionic transport properties of 
several distinct lithium FIC based on density functional theory AIMD simulations. Specifically, we analyze 
the lithium diffusivity and $T$-dependent density of vibrational states in the following compounds 
(space groups are indicated within parentheses): hexagonal Li$_{3}$N ($P6_{3}/mmc$) \cite{li10}, orthorhombic 
LiGaO$_{2}$ ($Pna2_{1}$) \cite{islam17}, cubic LiF ($Fm\overline{3}m$) \cite{yildirim15}, 
hexagonal LiIO$_{3}$ ($P6_{3}$) \cite{aliev88}, and tetragonal Li$_{3}$OCl ($P4/mmm$) \cite{zhao12}. 
Lattice phonons and activation energies for ion migration are calculated also at zero-temperature 
conditions to quantify the impact of temperature and anharmonic effects on their evaluation. Our 
simulation results demonstrate the lack of a direct correlation between $E_{a}$ and $\langle \omega \rangle$, 
where the latter term represents the average phonon frequency of the crystal (either associated to all 
the compound atoms or just Li ions). However, we show that the hopping attempt frequency of lithium ions, $\nu_{0}$, 
follows the Meyer-Neldel rule $\propto \exp{\left(E_{a}/\langle \omega \rangle\right)}$, in consistent 
agreement with recent experimental observations \cite{muy18b}. Thus, crystal anharmonicity, or equivalently
lattice softness, can be identified with enhanced ionic diffusivity but only within families of FIC that present 
inherently similar migration activation energies. On the technical side, we quantify the numerical inaccuracies 
in $E_{a}$ and $\langle \omega \rangle$ that result from neglecting temperature effects, which
unexpectedly turn out to be quite large (e.g., typical $\sim 10$\% underestimation of migration activation 
energies). Our theoretical work provides an improved understanding of how lattice dynamics affects lithium 
conductivity in FIC, hence it may be useful for improving the design of energy storage and energy conversion 
devices. Meanwhile, we substantiate the importance of considering temperature effects in first-principles 
modeling of lithium FIC.

\section{Simulation methods}
\label{sec:methods}

\subsection{First-principles calculations}
\label{subsec:DFT}
First-principles calculations based on density functional theory (DFT) are performed to analyse the vibrational 
and ionic transport properties of lithium FIC. We perform these calculations with the VASP code \cite{vasp} by 
following the generalized gradient approximation to the exchange-correlation energy due to Perdew \emph{et al.} 
\cite{pbe96}. (Possible dispersion interactions in Li$_{3}$N \cite{cazorla18} are captured with the D3 correction 
scheme developed by Grimme and co-workers \cite{grimmed3}.) The projector augmented-wave method is used to represent 
the ionic cores \cite{bloch94}, and the following electronic states are considered as valence: Li $1s$-$2s$, N 
$2s$-$2p$, Ga $4s$-$4p$, O $2s$-$2p$, F $2s$-$2p$, I $5s$-$5p$, and Cl $3s$-$3p$. Wave functions are represented 
in a plane-wave basis truncated at $650$~eV. By using these parameters and dense ${\bf k}$-point grids for Brillouin 
zone integration, the resulting energies are converged to within $1$~meV per formula unit. In the geometry relaxations,
a tolerance of $0.01$~eV$\cdot$\AA$^{-1}$ is imposed on the atomic forces. 

\emph{Ab initio} molecular dynamics (AIMD) simulations based on DFT are performed in the canonical $( N, V, T )$ 
ensemble (i.e., constant number of particles, volume, and temperature) for all the considered bulk materials. The 
selected volumes render zero-pressure conditions at room temperature, $T_{\rm room} = 300$~K. The temperature in the 
AIMD simulations is kept fluctuating around a set-point value by using Nose-Hoover thermostats. Large simulation 
boxes containing $N_{ion} \sim 250$ atoms are employed in all the cases, and periodic boundary conditions are 
applied along the three Cartesian directions. Newton's equations of motion are integrated by using the customary Verlet's 
algorithm and a time-step length of $\delta t = 10^{-3}$~ps. $\Gamma$-point sampling for integration within the first 
Brillouin zone is employed in all the AIMD simulations. The calculations comprise long simulation times of  
$t_{total} \sim 200$~ps. For each compound, we run a total of $8$ AIMD simulations at different temperatures and 
considering both stoichiometric and non-stoichiometric (that is, containing vacancies) systems. We focus on 
the description of the superionic and vibrational properties of lithium FIC, which are estimated by monitoring the 
positions and velocities of the ions during the AIMD simulations. Tests performed on the numerical bias stemming from the 
finite size of the simulation cell and duration of the molecular dynamics runs are reported in the Supplementary Methods. 
In view of the results obtained in such numerical tests, the adopted $N_{ion}$ and $t_{total}$ values can be assumed 
to provide reasonably well converged results for the ionic diffusivity and vibrational density of states of lithium FIC
(Supplementary Methods). 

Zero-temperature phonon frequency calculations are performed with the small-displacement method, in which 
the force-constant matrix is calculated in real-space by considering the proportionality between atomic 
displacements and forces \cite{cazorla17c,cazorla13,kresse95,alfe09}.
The quantities with respect to which our phonon calculations are converged include the size of the supercell,
the size of the atomic displacements, and the numerical accuracy in the sampling of the Brillouin zone. 
We find the following settings to provide quasi-harmonic free energies converged to within $5$~meV per
formula unit: $3 \times 3 \times 3$ supercells, typically containing $200$--$300$ ions (the figures indicate 
the number of replicas of the unit cell along the corresponding lattice vectors), atomic displacements of $0.02$~\AA, 
and ${\bf q}$-point grids of $14 \times 14 \times 14$. The value of the phonon frequencies are obtained with the 
PHON code developed by Alf\`e \cite{alfe09}. In using this code we exploit the translational invariance of the system, 
to impose the three acoustic branches to be exactly zero at the center of the Brillouin zone, and apply central 
differences in the atomic forces.

\emph{Ab initio} nudged-elastic band (NEB) calculations \cite{henkelman00} are performed to estimate the energy 
barriers for ionic diffusion in all investigated lithium FIC at zero temperature. Our NEB calculations typically 
are performed in $2 \times 2 \times 2$ or $3 \times 3 \times 3$ supercells containing several tens of atoms. We 
use ${\bf q}$-point grids of $8 \times 8 \times 8$ or $6 \times 6 \times 6$ and an energy plane-wave cut-off of 
$650$~eV. Six intermediate images are used to determine the most likely ionic diffusion paths when temperature 
effects are disregarded; the geometry optimizations are halted when the total forces on the atoms are smaller 
than $0.01$~eV$\cdot$\AA$^{-1}$.

\subsection{Estimation of key quantities}
\label{subsec:quantities}
The mean square displacement (MSD) is estimated with the formula:
\begin{eqnarray}
{\rm MSD}(\tau) & = & \frac{1}{N_{ion} \left( N_{step} - n_{\tau} \right)} \times \\ \nonumber
                &   & \sum_{i=1}^{N_{ion}} \sum_{j=1}^{N_{step} - n_{\tau}} | {\bf r}_{i} (t_{j} + \tau) - {\bf r}_{i} (t_{j}) |^{2}~, 
\label{eq1}
\end{eqnarray}
where ${\bf r}_{i}(t_{j})$ is the position of the migrating ion labelled as $i$ at time $t_{j}$ ($= j \cdot \delta t$), 
$\tau$ represents a lag time, $n_{\tau} = \tau / \delta t$, $N_{ion}$ is the total number of mobile ions, and $N_{step}$
the total number of time steps. The diffusion coefficient then is obtained by using the Einstein relation: 
\begin{equation}
D =  \lim_{\tau \to \infty} \frac{{\rm MSD}(\tau)}{6\tau}~.  
\label{eq2}	
\end{equation}
In practice, we consider $0 < \tau \le 100$~ps and estimate $D$ by performing linear fits over the last 
$\Delta \tau = 50$~ps.   

The $T$-dependence of the diffusion coefficient is assumed to follow the Arrhenius formula:
\begin{equation}
	D(T) = D_{0} \cdot \exp{\left[-\frac{E_{a}}{k_{B}T} \right]}~, 
\label{eq3}
\end{equation}
where $D_{0}$ is known as the pre-exponential factor, $E_{a}$ is the activation energy for ionic migration,
and $k_{B}$ the Boltzmann constant. 
From a physical point of view, $D_{0}$ can be interpreted as a hopping attempt frequency, $\nu_{0}$, the 
value of which is obtained via the relationship:
\begin{equation}
\nu_{0} = \frac{D_{0}}{a_{0}^{2}}~,
\label{eq4}	
\end{equation}
where $a_{0}$ represents the equilibrium lattice parameter of the crystal (that is, a characteristic 
length for the ionic hops). Likewise, the exponential factor in Eq.(\ref{eq3}) can be interpreted as an acceptance 
probability for the proposed ionic jumps. Hence, large (small) $D_{0}$ and small (large) $E_{a}$ lead to high (low) 
ionic conductivities. 

To estimate the density of vibrational states in lithium FIC, VDOS, we calculate the Fourier 
transform of the velocity-velocity autocorrelation function, directly obtained from the AIMD 
simulations, as:
\begin{equation}
	{\rm VDOS}(\omega) = \frac{1}{N_{ion}} \sum_{i}^{N_{ion}} \int_{0}^{\infty} 
	\langle {\bf v}_{i}(\tau)\cdot{\bf v}_{i}(0)\rangle e^{i\omega \tau} d\tau~,
\label{eq5}	
\end{equation}
where ${\bf v}_{i}(t)$ represents the velocity of the atom labelled as $i$ at time $t$, and $\langle \cdots \rangle$
denotes statistical average in the $( N, V, T )$ ensemble. We note that VDOS depends on temperature. 
Once the density of vibrational states is known, it is straightforward to calculate the corresponding phonon band center
or average lattice frequency, $\langle \omega \rangle$, defined as:
\begin{equation}
\langle \omega \rangle = \frac{\int_{0}^{\infty} {\rm VDOS}~ \omega~ d\omega}{\int_{0}^{\infty} {\rm VDOS}~ d\omega}~,
\label{eq6}
\end{equation}
which also depends on $T$. Likewise, the contribution of a particular group of ions to the full VDOS can be estimated 
by considering them alone in the summation appearing in Eq.(\ref{eq5}). In order to determine a characteristic low-energy 
phonon frequency for lithium FIC, we (somewhat arbitrarily) define the quantity:
\begin{equation}
	\langle \omega \rangle_{\rm room} = \frac{\int_{0}^{\omega_{\rm room}} {\rm VDOS}~ \omega~ d\omega}{\int_{0}^{\omega_{\rm room}} {\rm VDOS}~ d\omega}~,
\label{eq7}
\end{equation}
where $\omega_{\rm room}$ is $k_{B}T_{\rm room} / \hbar = 6.25$~THz.

\begin{figure*}
\centerline
        {\includegraphics[width=0.90\linewidth]{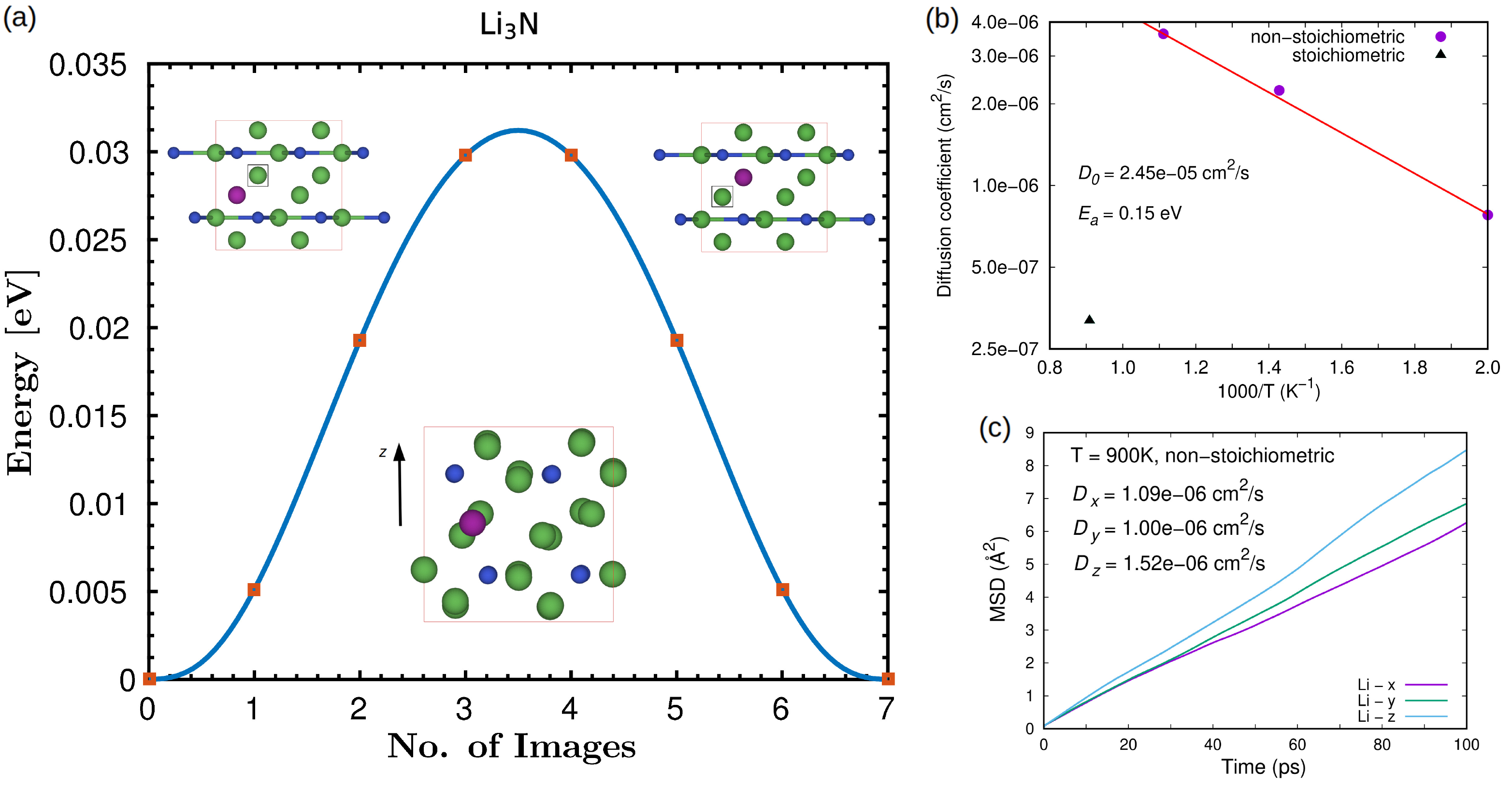}}
        \caption{Activation energy for ion migration calculated for Li$_{3}$N at (a)~$T = 0$~K and (b)~finite
        temperatures considering lithium vacancies. (c)~Ionic diffusivity estimated along the three Cartesian
        axis in non-stoichiometric Li$_{3}$N at $T = 900$~K. Lithium and nitrogen ions are represented with
        large green and small blue spheres, respectively. Lithium vacancy positions are indicated with black
        squares and mobile ions with purple spheres.}
\label{fig2}
\end{figure*}

\section{Results}
\label{sec:results}

\subsection{Li$_{3}$N}
\label{subsec:ln}

\begin{figure}
\centerline
        {\includegraphics[width=1.00\linewidth]{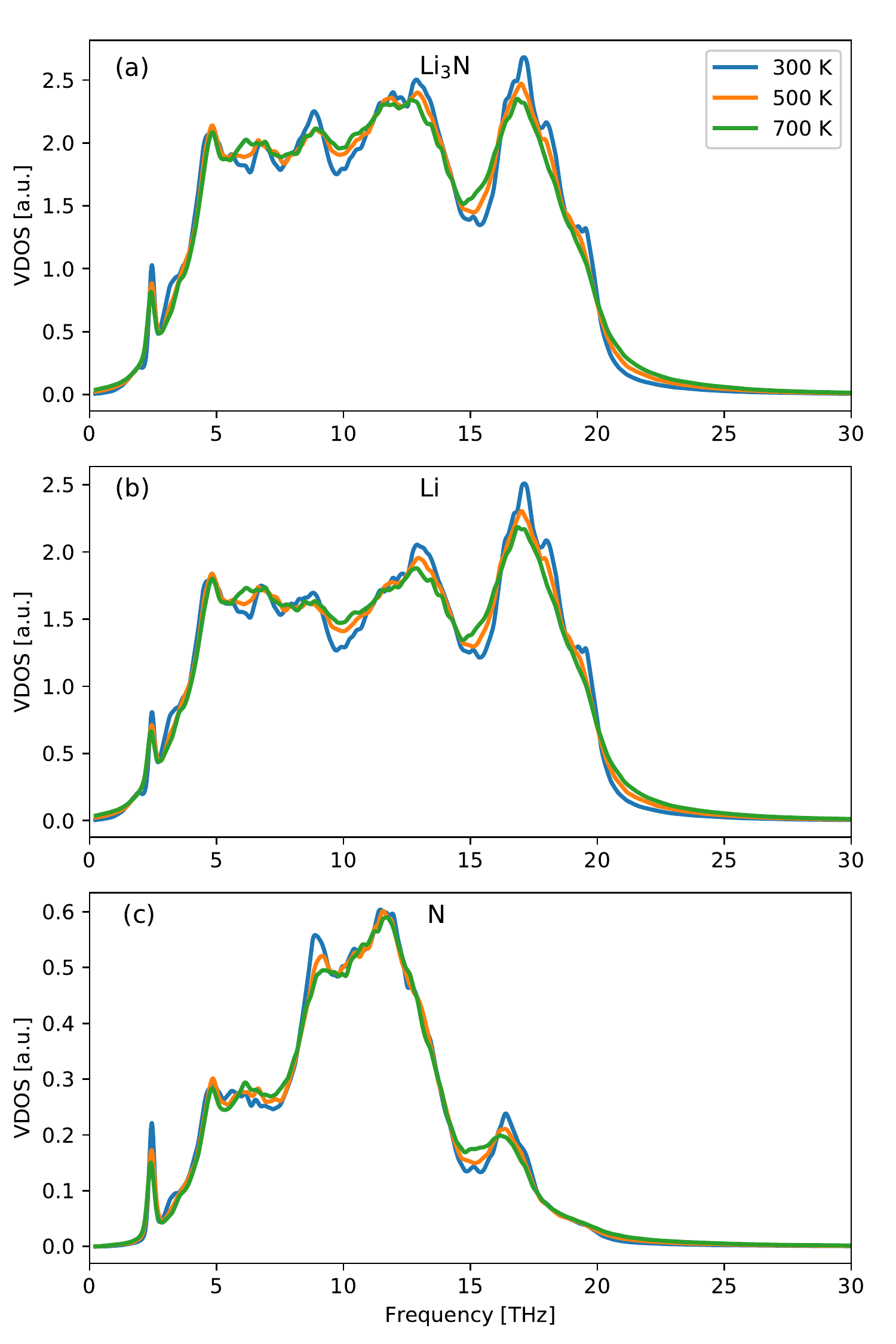}}
	\caption{Density of vibrational states calculated for Li$_{3}$N at different
	temperatures considering (a)~all the atoms, (b)~only Li ions, and (c)~only N ions. 
	Results are obtained from the Fourier transform of the velocity-velocity autocorrelation 
	function calculated during long AIMD simulations.}
\label{fig3}
\end{figure}

This compound presents two common polymorphs known as $\alpha$ and $\beta$ phases. The $\alpha$ phase 
(hexagonal, space group $P6/mmm$) has a layered structure composed of alternating planes of hexagonal 
Li$_{2}$N and pure Li$^{+}$ ions (see, for instance, Fig.1 in Ref.\cite{cazorla18}). The $\beta$ phase 
(hexagonal, space group $P6_{3}/mmc$) exhibits an additional layer of lithium ions intercalated between 
consecutive Li$_{2}$N planes that is accompanied by a doubling of the unit cell (Fig.\ref{fig2}a). Exceptionally 
high ionic conductivities of the order of $10^{-4}$--$10^{-3}$~S~cm$^{-1}$ have been measured experimentally 
in Li$_{3}$N at room temperature \cite{li10,alpen77,nazri94}. Here, we restrict our analysis to $\beta$--Li$_{3}$N. 

Figure~\ref{fig2}a shows the minimum ion migration activation energy calculated for Li$_{3}$N at zero temperature 
with the NEB method, $E_{a}^{0}$. The local minimum-energy points correspond to lithium vacancy positions, since we 
were not able to generate any metastable interstitial configuration for the stoichiometric system. This outcome 
suggests that superionicity in Li$_{3}$N is vacancy mediated, hence mostly it occurs in non-stoichiometric systems. 
In good agreement with previous zero-temperature DFT results reported by Li \emph{et al.} \cite{li10}, we find that 
the minimum $E_{a}^{0}$ amounts to $0.03$~eV and corresponds to lithium diffusion in the region contained between 
consecutive Li$_{2}$N planes lying along the $z$ direction (Fig.\ref{fig2}a). Thus, ionic diffusion seems to be mostly 
confined to two dimensions, which we denote here as $x$-$y$. We note that the migration activation energy determined 
in experiments, $E_{a}^{\rm expt}$, is $0.45$~eV \cite{li10}, which is significantly larger than $E_{a}^{0}$. 

Figures~\ref{fig2}b and c show the results of our AIMD simulations performed for Li$_{3}$N at finite temperatures.
In accordance with the zero-temperature results just explained, lithium conductivity appears to be vacancy mediated
since the diffusion coefficients calculated in the stoichiometric system at high temperatures are too small 
(namely, $D < 10^{-6}$~cm$^{2}$s$^{-1}$ at $T > 1000$~K, Fig.~\ref{fig2}b). However, the lithium diffusion mechanisms 
and activation energy that are deduced from the $T \neq 0$ simulations differ appreciably from those obtained with 
zero-temperature methods. In particular, the lithium ions are found to diffuse almost equally along all three Cartesian 
directions (Fig.~\ref{fig2}c), and the estimated activation energy is much larger, $E_{a} = 0.15$~eV (Fig.~\ref{fig2}b).
Although the agreement between the estimated and experimentally measured activation energies has been improved, there 
is still a considerable difference between $E_{a}$ and $E_{a}^{\rm expt}$. A likely explanation for such a discrepancy 
may be the neglection of defects other than vacancies in the AIMD simulations (e.g., cracks, dislocations, and interfaces), 
which in some cases are known to deplete ionic diffusivity significantly \cite{sun15,adepalli17}. Another possibility 
is that the concentration of extrinsic vacancies in our AIMD simulations probably is too large (that is, $\sim 2$\%), 
which may lead to an overestimation of lithium conductivity. Meanwhile, we estimate a pre-exponential factor of 
$D_{0} = 2.5 \cdot 10^{-5}$~cm$^{2}$s$^{-1}$ and a hopping attempt frequency, $\nu_{0}$, of $\sim 0.01$~THz. 

Figure~\ref{fig3}a shows the density of vibrational states, VDOS, estimated for non-stoichiometric Li$_{3}$N 
at several temperatures with AIMD methods. The system is vibrationally stable in all the cases since no imaginary 
phonon frequencies appear in the corresponding phonon spectra. This outcome is contrary to the results obtained with 
the harmonic approximation at zero temperature, which indicate that non-stoichiometric Li$_{3}$N is vibrationally 
unstable (see Fig.~\ref{fig1}); we note that the imaginary eigenfrequency phonon modes appearing in that 
zero-temperature harmonic VDOS are mostly dominated by Li ions (see Supplementary Fig.2). As the temperature 
is increased, the peaks of the VDOS become smoothed and the resulting phonon band center, $\langle \omega \rangle$, 
increases steadily (due to the fact that higher-frequency vibrational modes become thermally activated). The accompanying 
increase in $\langle \omega \rangle$, however, is very mild. For instance, the phonon band center amounts to $11.66$~THz 
at room temperature and to $11.74$~THz at $T = 700$~K. The average phonon frequency that is estimated by applying a 
cut-off of $k_{B}T_{\rm room} / \hbar$ to the VDOS, $\langle \omega \rangle_{\rm room}$ (Eq.(\ref{eq7}) in 
Sec.~\ref{subsec:quantities}), provides a characteristic frequency for the low-energy phonons of the material, 
which are mostly related to the mechanical stiffness of the material. We find that the $T$-induced variation of 
such a frequency is also very moderate; for example, at $T = 700$~K $\langle \omega \rangle_{\rm room}$ 
is $\sim 1$\% larger than the value estimated at ambient conditions, which is $4.7$~THz. 

We note that at $T = 700$~K non-stoichiometric Li$_{3}$N is superionic and presents a large diffusion 
coefficient of $2.2 \cdot 10^{-6}$~cm$^{2}$s$^{-1}$ whereas at room temperature remains in the normal 
state ($D \sim 0$). Hence, for a same lithium FIC it seems not possible to correlate the large $T$-induced 
variations in ionic conductivity with the accompanying changes in VDOS, which are minute (i.e., of the order 
of $\sim 0.1$~THz). Figures~\ref{fig3}b and c show the partial VDOS corresponding to lithium and nitrogen ions, 
respectively. The partial phonon band center of the nitrogen ions is lower than $\langle \omega \rangle_{\rm Li}$ 
by $\sim 3$\%, as it could have been foreseen due to their larger atomic mass (recall the $m_{\alpha}^{-1/2}$ 
factors entering the expression of the dynamical force constant matrix \cite{cazorla17c,cazorla13,cazorla17d}). 
Interestingly, low anion vibration phonon excitations have been linked to a reduction in FIC stability against 
electrochemical oxidation by Muy \emph{et al.} \cite{muy18}.

\subsection{LiGaO$_{2}$}
\label{subsec:lgo}

\begin{figure*}
\centerline
        {\includegraphics[width=0.90\linewidth]{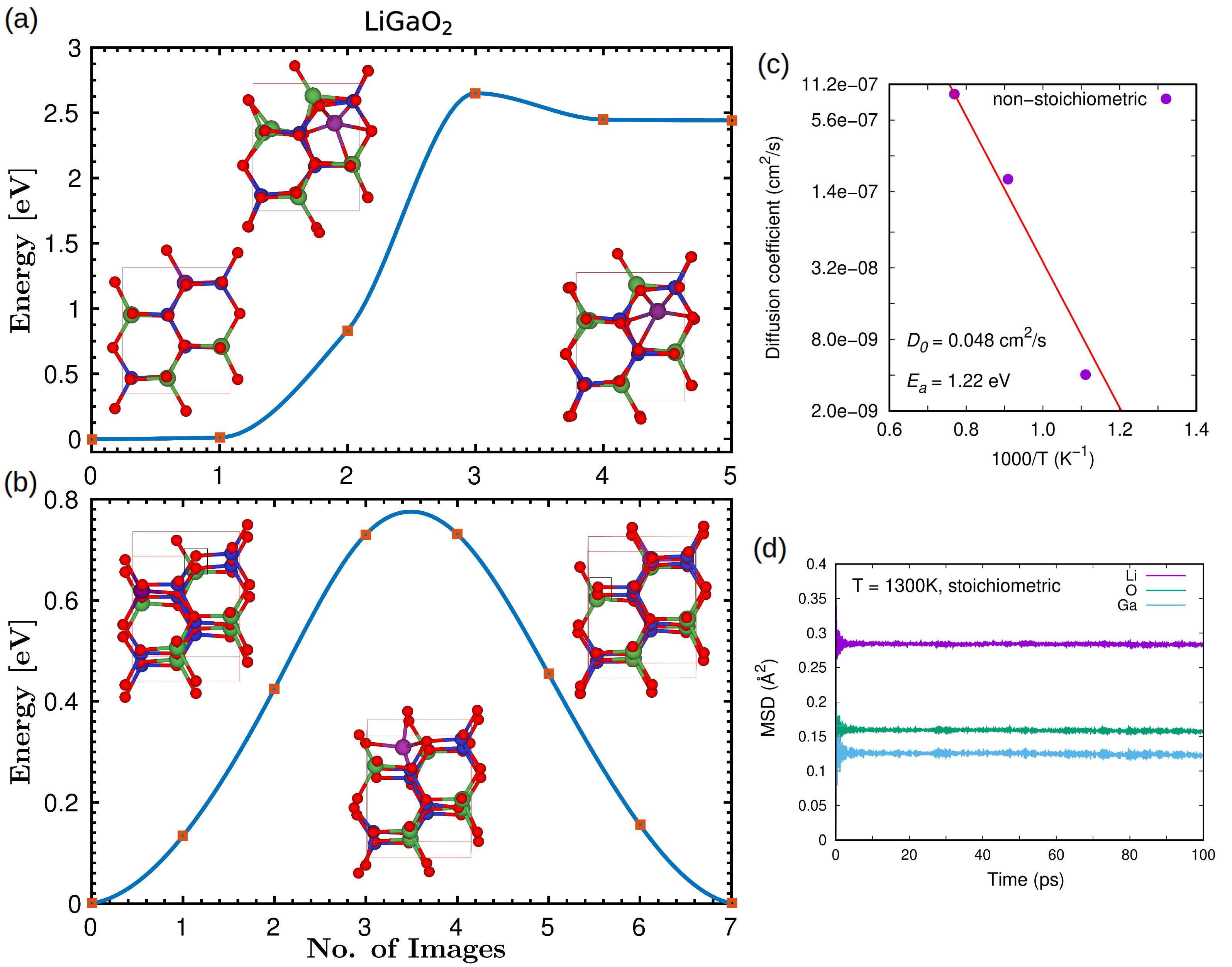}}
\caption{Activation energy for ion migration calculated for LiGaO$_{2}$ at zero-temperature considering 
	(a)~interstial and (b)~vacancy positions. (c)~Lithium diffusivity of non-stoichiometric LiGaO$_{2}$ 
	calculated at finite temperatures. (d)~Mean square displacement of stoichiometric LiGaO$_{2}$ calculated 
	at very high temperatures. Lithium, gallium, and oxygen ions are represented with large green, 
	small blue, and small red spheres, respectively. Lithium vacancy positions are indicated with 
	black squares and mobile ions with purple spheres.}
\label{fig4}
\end{figure*}

At ambient conditions this compound stabilizes in a orthorhombic structure with space group $Pna2_{1}$.
The lithium and gallium ions are located at the center of oxygen tetrahedrons, forming a two-dimensional
stacking of alternating LiO$_{4}$ and GaO$_{4}$ arrays. Recently, a combined experimental and theoretical
study has proved the superionic nature of LiGaO$_{2}$ at temperatures $T \ge 800$~K \cite{islam17}.

Figures~\ref{fig4}a and b show the activation energy for Li interstitial and vacancy migration calculated 
at zero temperature with the NEB method. In this case, we were able to generate a metastable interstitial 
configuration for the stoichiometric system, as shown in Fig.\ref{fig4}a; however, the accompanying 
zero-temperature interstitial formation energy and activation migration energy appear to be too large, 
namely, $2.4$ and $2.6$~eV respectively. Meanwhile, the estimated zero-temperature activation energy for
vacancy migration is considerably lower, $E_{a}^{0} = 0.78$~eV (Fig.\ref{fig4}b). These results suggest that 
fast-ion conductivity in LiGaO$_{2}$ is vacancy mediated, which is in agreement with previous NEB DFT 
calculations reported by Islam \emph{et al.} for the same system \cite{islam17}. We note, however, that the 
experimentally measured activation energy for Li migration is $E_{a}^{\rm expt} = 1.25$~eV \cite{islam17}, 
which is significantly larger than the predicted $E_{a}^{0}$. A possible explanation for the significant 
discrepancy between the measured and calculated zero-temperature activation energies could be, among other 
causes, the neglection of temperature effects in NEB simulations, as it has been suggested by the authors 
of Ref.\cite{islam17}.

Our AIMD results shown in Fig.\ref{fig4}c confirm that after considering temperature effects the agreement 
between the estimated and measured activation energies improves drastically. In particular, we calculate 
a large $E_{a}$ of $1.22$~eV, which practically coincides with the corresponding experimental value. Likewise, 
the computed pre-exponential factor is $5.0 \cdot 10^{-2}$~cm$^{2}$s$^{-1}$, which leads to a very large $\nu_{0}$ 
of $\sim 10$~THz. In agreement with the zero-temperature results explained above, lithium conductivity in LiGaO$_{2}$ 
appears to be vacancy mediated since even at high temperatures of $T > 1000$~K the stoichiometric system remains 
in the normal state (Fig.\ref{fig4}d).

We have estimated the VDOS of non-stoichiometric LiGaO$_{2}$ at several temperatures with AIMD methods (Supplementary
Fig.3). The system is vibrationally stable in all the cases since no imaginary phonon frequencies appear in the 
corresponding phonon spectra. The effect of temperature on the calculated VDOS is very small and similar to that 
described previously for Li$_{3}$N. For instance, the average phonon frequency $\langle \omega \rangle$ 
amounts to $11.74$~THz at $T = 900$~K and to $11.83$~THz at $1300$~K. Thus, again the large diffusion 
coefficient changes induced by temperature ($\Delta D / D \sim 10^{2}$ for $\Delta T = 400$~K) 
do not appear to be reflected on the corresponding VDOS ($\Delta \langle \omega \rangle / \langle \omega 
\rangle \sim 10^{-2}$ for the same $\Delta T$). An analagous increase of about $1$\% is 
found for $\langle \omega \rangle_{\rm room}$, which at $T = 900$~K amounts to $4.4$~THz. Regarding the 
partial VDOS, we find that $\langle \omega \rangle_{\rm Ga} = 0.6 \langle \omega \rangle_{\rm Li}$ and $\langle \omega 
\rangle_{\rm O} = 1.2 \langle \omega \rangle_{\rm Li}$, thus in average the Ga ions vibrate at lower frequencies
than the lithium ions while the oxygen atoms at higher. Despite the larger mass of the oxygen atoms as compared 
to lithium, $\langle \omega \rangle_{\rm O}$ is the highest because the oxygen atoms participate in all  
covalent bonds of the crystal and therefore appear represented along the whole VDOS (Supplementary Fig.3). 
Consequently, the presence of oxygens in Li FIC will tend to increase the phonon band center of the anion 
sublattice; this effect, however, is likely to be reduced significantly in the presence of other electronegative 
species with larger atomic masses (e.g., sulfur) \cite{muy18}.

\subsection{LiF}
\label{subsec:lf}

\begin{figure*}
\centerline
        {\includegraphics[width=0.90\linewidth]{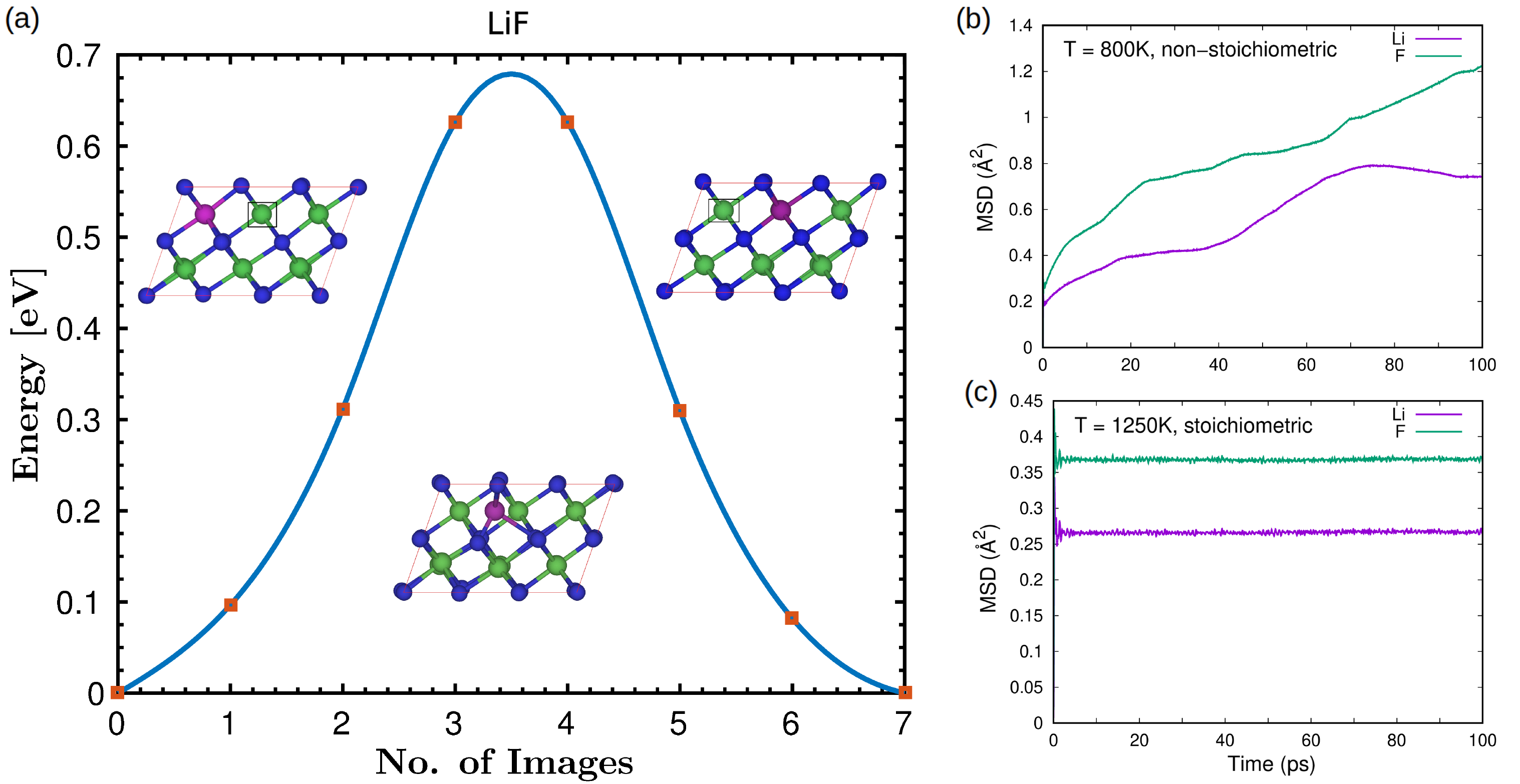}}
	\caption{(a)~Activation energy for ion migration calculated for LiF at zero-temperature considering
        vacancy positions. (b)~Mean square displacement of non-stoichiometric LiF calculated at
        high temperatures. (c)~Mean square displacement of stoichiometric LiF calculated at very
        high temperatures. Lithium and fluorine ions are represented with large green and small 
	blue spheres, respectively. Lithium vacancy positions are indicated with black squares 
	and mobile ions with purple spheres.}
\label{fig5}
\end{figure*}

The crystal structure of bulk LiF at ambient conditions is rocksalt (cubic, space group $Fm\overline{3}m$). 
The presence of LiF has been revealed in the interface formed between the solid electrolyte and electrodes 
in Li-ion batteries \cite{nie13}, and the accompanying effects on energy storage performance have been 
investigated recently by several authors with theory \cite{yildirim15,pan15,soto18}. 

By performing NEB calculations, we estimate a zero-temperature energy barrier for lithium vacancy diffusion 
of $E_{a}^{0} = 0.68$~eV (Fig.\ref{fig5}a), which is in very good agreement with previous DFT calculations 
\cite{yildirim15,pan15,soto18}. We note that in this case we could neither find a metastable interstitial 
configuration, hence ionic conductivity in LiF in principle appears to be vacancy mediated. The $E_{a}^{0}$ 
calculated for LiF is comparable to that determined previously for LiGaO$_{2}$, suggesting that lithium 
diffusivity in both materials should be similar. However, our AIMD simulations show this not to be the case. 
In particular, at temperatures above $700$~K non-stoichiometric LiF becomes vibrationally unstable, as shown 
by the fact that both the lithium and fluorine ions become mobile (Fig.\ref{fig5}b). Consequently, we have 
not been able to determine any $E_{a}$ or $D_{0}$ for bulk LiF. This outcome highlights the importance of atomic 
structure on lithium conductivity: two materials with similar $E_{a}^{0}$ but different geometries may be not 
comparable in terms of ionic diffusivity. Meanwhile, our AIMD simulations show that lithium conduction in 
stoichiometric LiF is negligible even at high temperatures (that is, $D \sim 0$ at $T > 1000$~K, Fig.\ref{fig5}c), 
in agreement with the zero-temperature NEB calculations. 

The $T$-dependent VDOS of non-stoichiometric LiF calculated with AIMD simulations are reported in Supplementary 
Fig.4. The corresponding average phonon frequency amounts to $9.19$~THz at $T = 400$~K and to $9.45$~THz at $600$~K. 
Hence, despite the fact that LiF is a much worse lithium conductor than, for instance, Li$_{3}$N, its phonon band 
center is significantly lower (e.g., $\langle \omega \rangle = 11.70$~THz at $T = 500$~K for non-stoichiometric 
Li$_{3}$N). Such a comparison suggests the lack of a direct correlation between lattice dynamics and lithium 
diffusivity in the investigated systems; we will comment on this point with more detail later on. Regarding the 
partial VDOS, we find that $\langle \omega \rangle_{\rm F} = 0.7 \langle \omega \rangle_{\rm Li}$ owing to the larger 
mass of the fluorine ions and the ionic nature of the material. Thus, the low-frequency lattice excitations in LiF 
($\omega < 5$~THz) are dominated by the fluorine ions, rather than by Li.

\subsection{LiIO$_{3}$}
\label{subsec:lio}

\begin{figure*}
\centerline
        {\includegraphics[width=0.90\linewidth]{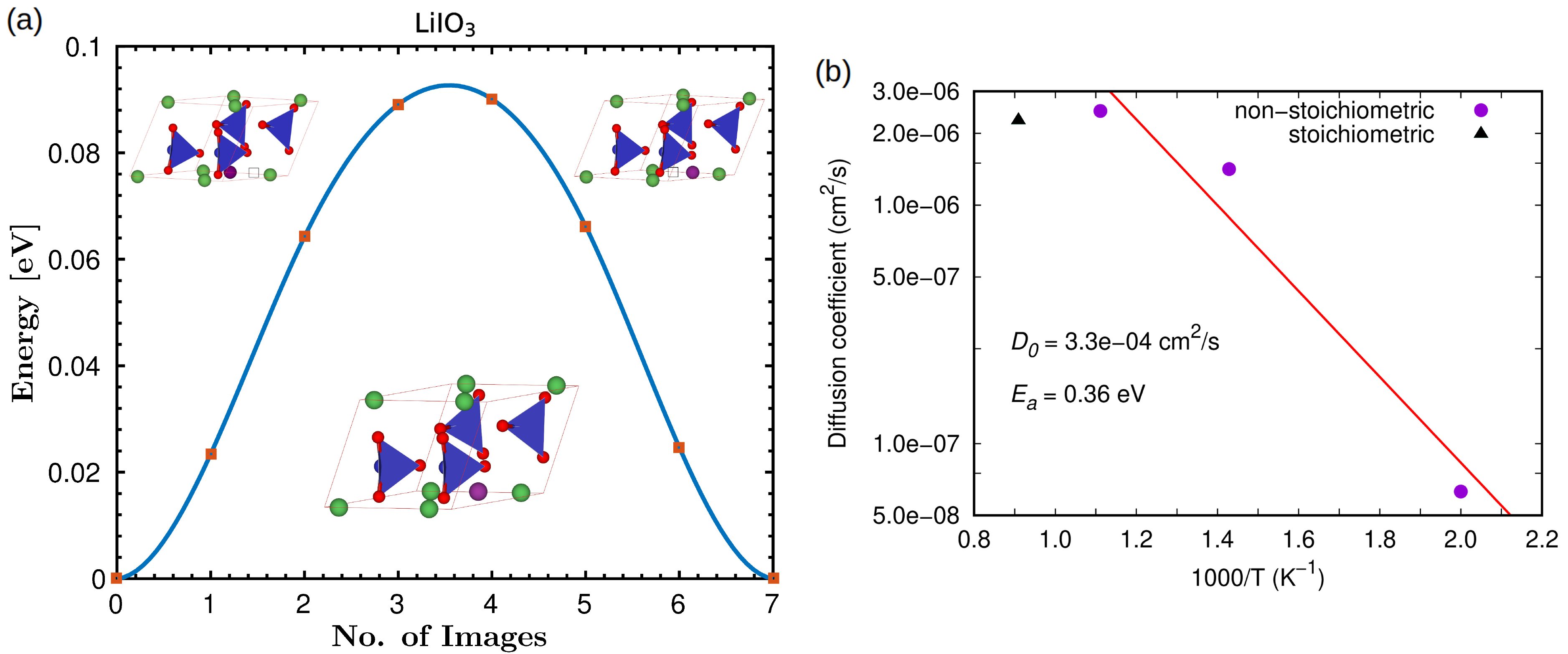}}
	\caption{(a)~Activation energy for ion migration calculated for LiIO$_{3}$ at zero-temperature 
	considering vacancy positions. (b)~Lithium diffusivity of non-stoichiometric and stoichiometric
	LiIO$_{3}$ calculated at finite temperatures. Lithium, iodine, and oxygen ions are represented 
	with large green, small blue, and small red spheres, respectively. Lithium vacancy positions 
	are indicated with black squares and mobile ions with purple spheres.}
\label{fig6}
\end{figure*}

At ambient conditions this compound stabilizes in a hexagonal phase (space group $P6_{3}$) in which each
iodine atom is surrounded by three oxygen atoms, forming a three-dimensional net of tightly bound pyramidal 
IO$_{3}^{-}$ groups (Fig.\ref{fig6}a). LiIO$_{3}$ has a very complex polymorphism and undergoes reconstructive 
phase transitions by effect of pressure and temperature \cite{liang89,zhang02}. More than four decades ago, 
Aliev \emph{et al.} experimentally investigated the mobility of lithium ions in this compound \cite{aliev88}. 
They concluded that ionic diffusion occurred within quasi one-dimensional channels oriented along the
hexagonal $c$-axis, with a corresponding activation energy of $E_{a}^{\rm expt} = 0.26$~eV \cite{aliev88}. 
To the best of our knowledge, the lithium diffusion mechanisms in LiIO$_{3}$ have not been studied
previously with first-principles methods.

By using NEB calculations, we estimate a zero-temperature activation energy of $E_{a}^{0} = 0.09$~eV for 
lithium vacancies diffusing along the hexagonal $c$ direction (Fig.\ref{fig6}a). This result is significantly 
smaller than the corresponding experimental value. Nevertheless, our AIMD simulations render $E_{a} = 0.36$~eV
(Fig.\ref{fig6}b), which is larger than $E_{a}^{0}$ and provides a better agreement with the experiments. 
Likewise, the calculated pre-exponential factor is $3.5 \cdot 10^{-4}$~cm$^{2}$s$^{-1}$ and the resulting 
hopping attempt frequency $\sim 0.1$~THz. Large diffusion coefficients are estimated also for stoichiometric 
LiIO$_{3}$, although at temperatures well above ambient conditions (Fig.\ref{fig6}b). 

The VDOS calculated for non-stoichiometric LiIO$_{3}$ at $T \neq 0$ conditions are reported in Supplementary 
Fig.5. The total phonon band center amounts to $9.82$~THz at room temperature and increases to $9.92$~THz at 
$T = 700$~K. Once again, the huge changes induced by temperature on the diffusion coefficient ($\Delta D / D \sim 
10^{3}$ for $\Delta T = 400$~K) are not reflected on the corresponding VDOS ($\Delta \langle \omega \rangle / 
\langle \omega \rangle \sim 10^{-2}$ for the same $\Delta T$). We also note that although the $E_{a}$ estimated 
for LiIO$_{3}$ is about two times larger than estimated for Li$_{3}$N, the average phonon frequency in the former 
compound is noticeably smaller (namely, $9.82$ and $11.66$~THz, respectively, at $T_{\rm room}$). In analogy to 
LiGaO$_{2}$, we find that the oxygen ions render the largest $\langle \omega \rangle$ whereas the heaviest cations 
the smallest ($\langle \omega \rangle_{\rm I} = 0.4 \langle \omega \rangle_{\rm Li}$); in this case, the low-frequency 
lattice excitations ($\omega < 5$~THz) are clearly dominated by oxygen and iodine ions (Supplementary Fig.5).

\subsection{Li$_{3}$OCl}
\label{subsec:lclo}

\begin{figure*}
\centerline
        {\includegraphics[width=0.90\linewidth]{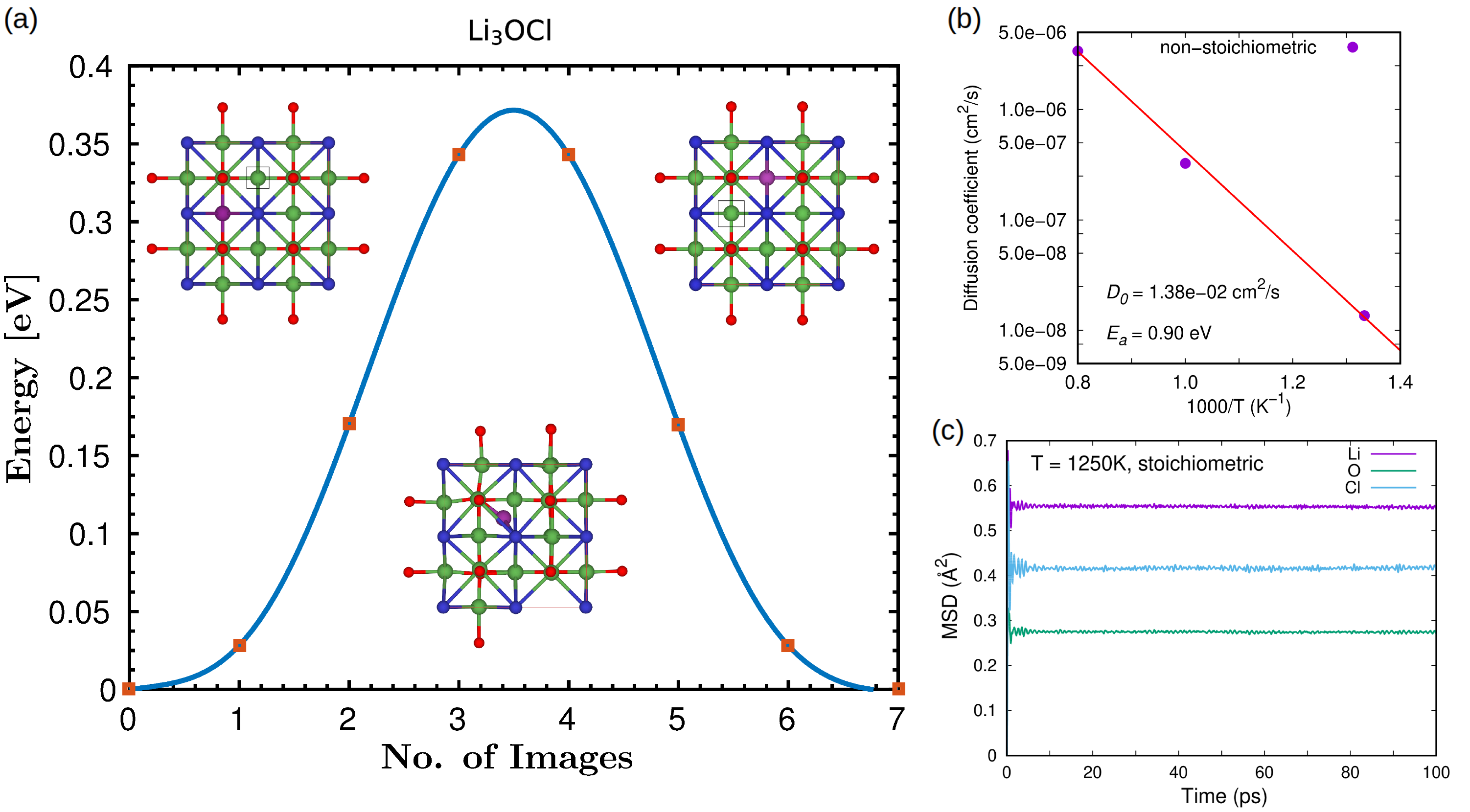}}
	\caption{(a)~Activation energy for ion migration calculated for Li$_{3}$OCl at zero-temperature 
	considering vacancy positions. (b)~Lithium diffusivity of non-stoichiometric Li$_{3}$OCl calculated 
	at finite temperatures. (c)~Mean square displacement of stoichiometric Li$_{3}$OCl calculated at
        very high temperatures. Lithium, chlorine, and oxygen ions are represented with large green,
        small blue, and small red spheres, respectively. Lithium vacancy positions are indicated with
        black squares and mobile ions with purple spheres.}
\label{fig7}
\end{figure*}

\begin{table*}
\centering
\begin{tabular}{c c c c c c c}
\hline
\hline
$ $ & $ $ & $ $ & $ $ & $ $ & $ $ & $ $ \\
	$ $ \qquad & \qquad  $E_{a}^{0}$ \qquad & \qquad $E_{a}$ \qquad & \qquad \qquad $E_{a}^{\rm expt}$ \qquad \qquad & \quad \qquad $D_{0}$ \quad \qquad & \qquad $\nu_{0}$ \qquad & \qquad $\langle \omega \rangle_{\rm room}$ \qquad \\
${\rm Material} $ \qquad & \qquad ${\rm (eV)}$ \qquad & \qquad ${\rm (eV)}$ \qquad & \qquad ${\rm (eV)}$ \qquad & \qquad ${\rm (cm^{2}/s)}$ \qquad & \qquad ${\rm (s^{-1})}$ \qquad & \qquad ${\rm (s^{-1})}$ \qquad \\
$ $ & $ $ & $ $ & $ $ & $ $ & $ $ & $ $ \\
\hline
$ $ & $ $ & $ $ & $ $ & $ $ & $ $ & $ $ \\
	${\rm Li_{3}N}$ \qquad   & $0.03$ & $0.15 \pm 0.01$ & $0.45$~\cite{li10}   & $2.5 \pm 1.0 \cdot 10^{-5}$ & $\sim 10^{10}$ & $4.7 \cdot 10^{12}$ \\
$ $ & $ $ & $ $ & $ $ & $ $ & $ $ & $ $ \\
	${\rm LiGaO_{2}}$ \qquad & $0.78$ & $1.2 \pm 0.1 $ & $1.25$~\cite{islam17} & $5.0 \pm 1.5 \cdot 10^{-2}$ & $\sim 10^{13}$ & $4.4 \cdot 10^{12}$ \\
$ $ & $ $ & $ $ & $ $ & $ $ & $ $ & $ $ \\
	${\rm LiF}$ \qquad       & $0.68$ & $ - $  & $ - $     & $ - $                & $ - $          & $5.0 \cdot 10^{12}$ \\
$ $ & $ $ & $ $ & $ $ & $ $ & $ $ & $ $ \\
	${\rm LiIO_{3}}$ \qquad  & $0.09$ & $0.36 \pm 0.07$ & $0.26$~\cite{aliev88}    & $3.5 \pm 1.5 \cdot 10^{-4}$ & $\sim 10^{11}$ & $4.2 \cdot 10^{12}$ \\
$ $ & $ $ & $ $ & $ $ & $ $ & $ $ & $ $ \\
	${\rm Li_{3}OCl}$ \qquad & $0.37$ & $0.90 \pm 0.05$ & $0.26$~\cite{zhao12}    & $1.4 \pm 0.5 \cdot 10^{-2}$ & $\sim 10^{13}$ & $4.4 \cdot 10^{12}$ \\
$ $ & $ $ & $ $ & $ $ & $ $ & $ $ & $ $ \\
\hline
\hline
\end{tabular}
\label{tab:summary}
\caption{Summary of the activation energies for ion migration calculated at zero and finite temperatures for the different
	 lithium FIC considered in this study. Experimental values for $E_{a}$ are reported for comparison purposes. The 
	 accompanying pre-exponential factors of lithium diffusivity, $D_{0}$, and corresponding hopping attempt frequencies, 
	 $\nu_{0}$, are enclosed along with the energy scale of the lattice phonon excitations calculated at room temperature, 
	 $\langle \omega \rangle_{\rm room}$ (estimated by using a frequency cut-off of $k_{B}T_{\rm room}/\hbar$ 
	 on the total density of vibrational states).}
\end{table*}

This compound presents an anti-perovskite structure similar to that of archetypal ABO$_{3}$ perovskite oxides 
\cite{zhao12,cazorla15,cazorla14b}. Specifically, the Li, O, and Cl ions are placed at octahedral vertices, octahedral 
centers, and cube vertices, respectively. (We note that in our zero-temperature geometry relaxations the  
Li$_{3}$OCl unit cell presents a $c/a$ ratio of $0.97$; hence, our symmetry labelling as tetragonal 
$P4/mmm$ rather than as cubic $Pm\overline{3}m$.) The main mechanism for ion migration in Li$_{3}$OCl has been 
proposed to be vacancy diffusion accompanied by anion disorder \cite{zhao12,zhang13,lu15}. 

Our zero-temperature NEB calculations render $E_{a}^{0} = 0.37$~eV for lithium vacancy diffusion (Fig.\ref{fig7}a), 
which is in good agreement with previous first-principles results \cite{zhang13,lu15} and the experimental 
value $E_{a}^{\rm expt} = 0.26$~eV \cite{zhao12}. Our AIMD simulations, however, provide a much larger value of the 
lithium migration enery barrier, namely, $E_{a} = 0.90$~eV (Fig.\ref{fig7}b). The corresponding pre-exponential factor 
is very large as well, $D_{0} = 1.4 \cdot 10^{-2}$~cm$^{2}$s$^{-1}$, which leads to a high hopping attempt frequency of 
$\sim 10$~THz (very similar to the one estimated previously for LiGaO$_{2}$). A possible cause for the large discrepancy 
between $E_{a}^{\rm expt}$ and $E_{a}$ may be the presence of a higher concentration of anion disorder and lithium 
vacancies in the experimental samples (in our AIMD simulations we have considered only vacancies at a small concentration 
of $\sim 1$\%). Meanwhile, our AIMD simulations confirm that lithium diffusivity in stoichiometric Li$_{3}$OCl is negligible
\cite{zhang13,lu15}, even at high temperatures of $T > 1000$~K (Fig.\ref{fig7}c). 

The VDOS calculated for non-stoichiometric Li$_{3}$OCl considering anharmonic and temperature effects are reported 
in Supplementary Fig.6. The total phonon band center amounts to $8.90$~THz at $T = 1000$~K and increases to $9.08$~THz 
at $T = 1250$~K. Like in the previous cases, the huge changes induced by temperature on the diffusion coefficient ($\Delta 
D / D \sim 10^{1}$ for $\Delta T = 250$~K) do not translate into significant VDOS variations ($\Delta \langle \omega 
\rangle / \langle \omega \rangle \sim 10^{-2}$ for the same $\Delta T$). We also note that although the $E_{a}$ 
calculated for Li$_{3}$ClO is about three times larger than estimated for LiIO$_{3}$, the total average phonon 
frequency of both compounds are quite similar ($8.9$ and $9.9$~THz, respectively, at $T \sim 1000$~K). In analogy to 
previous cases, we find that the oxygen ions render the largest $\langle \omega \rangle$ whereas the heaviest cations the 
smallest ($\langle \omega \rangle_{\rm Cl} = 0.5 \langle \omega \rangle_{\rm Li}$). On the other hand, the low-frequency 
lattice excitations ($\omega < 5$~THz) are dominated by lithium and chlorine ions (Supplementary Fig.6).

\begin{figure}
\centerline
	{\includegraphics[width=1.0\linewidth]{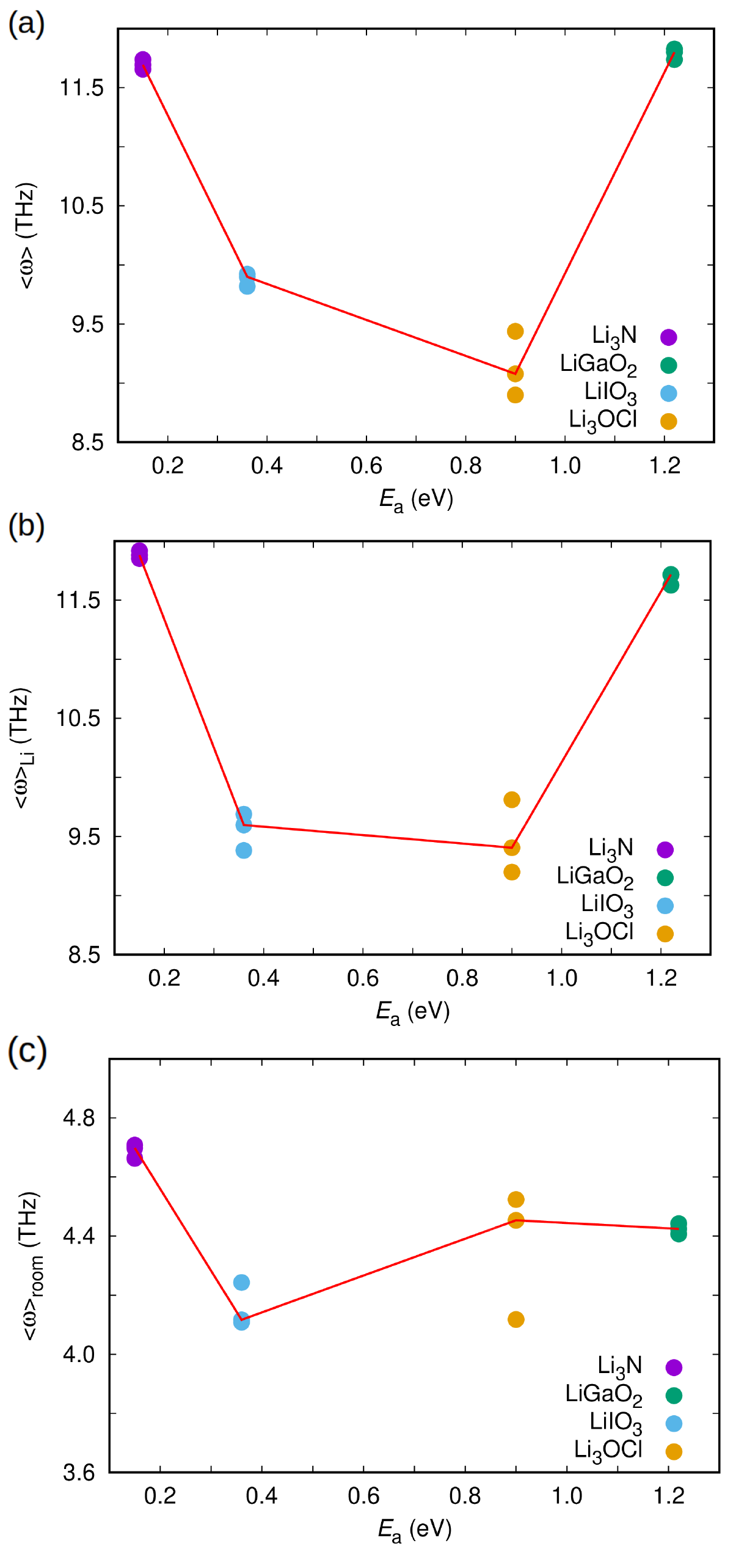}}
	\caption{Phonon band center versus the activation energy barrier for ion 
	migration calculated in lithium FIC considering (a)~all the 
	atoms in the crystal, (b)~only the Li ions, and (c)~all the atoms in the crystal 
	and a frequency cut-off of $k_{B}T_{\rm room}/\hbar$. Solid lines are guides to the eye.}
\label{fig8}
\end{figure}

\begin{figure}
\centerline
        {\includegraphics[width=1.0\linewidth]{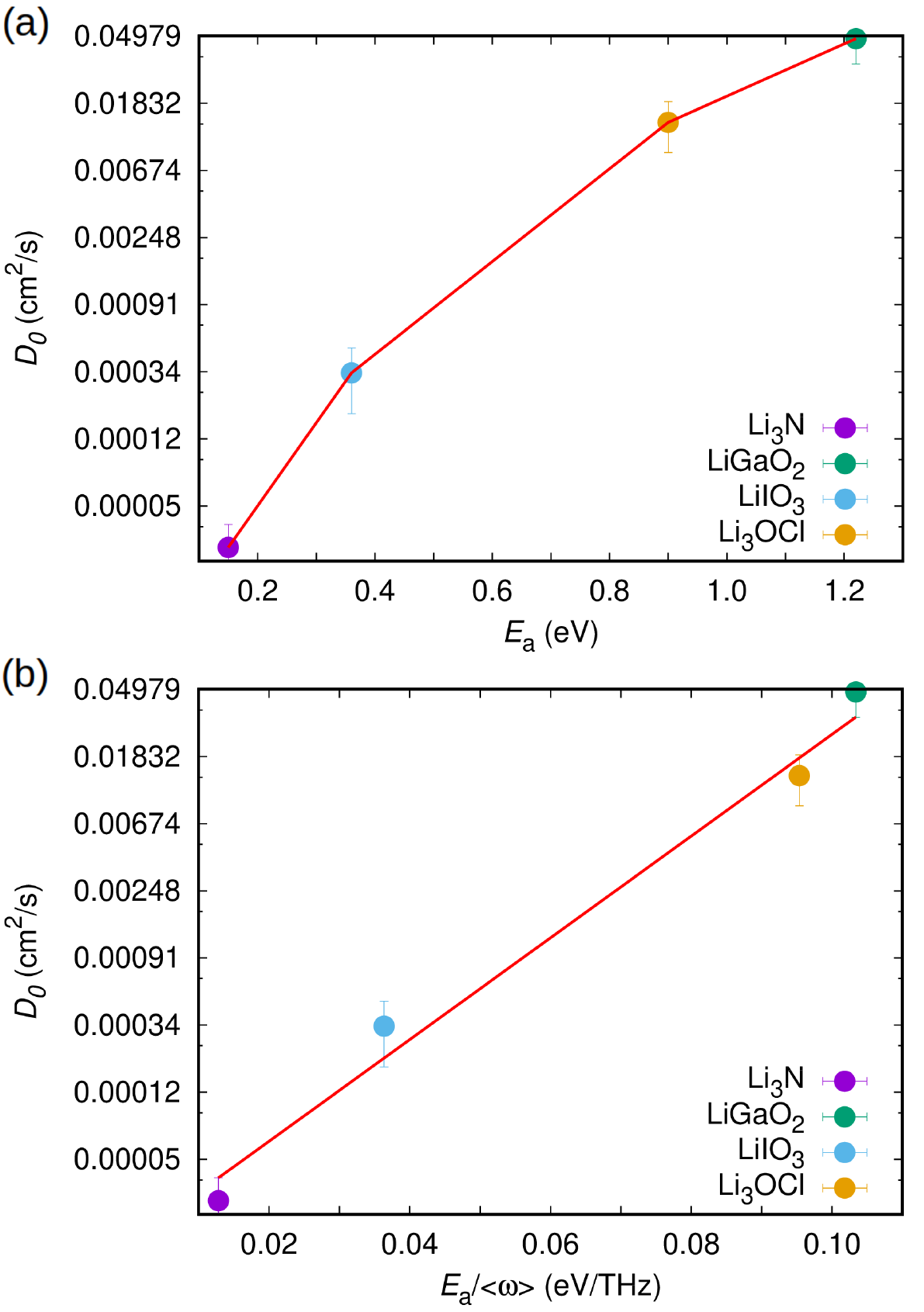}}
        \caption{Lithium diffusivity pre-exponential factor (in logarithmic scale) versus 
        the (a)~activation energy barrier for ion migration (the solid line is a guide to the
        eye), and (b)~ratio of the activation energy barrier for ion migration and phonon 
        band center (the solid line corresponds to a linear fit).}
\label{fig9}
\end{figure}

\section{Discussion}
\label{subsec:discussion}
Table~1 encloses a summary of the main results obtained for the five lithium FIC analysed in this
study. Next, we comment on (i)~the importance of considering temperature effects on the estimation of migration
activation energies and phonon frequencies in lithium FIC, and (ii)~the correlations between superionic descriptors
and lattice dynamics that can be deduced from our AIMD simulations.

\subsection{Temperature effects on $E_{a}$ and $\langle \omega \rangle$}
\label{subsec:teffects}
The differences between columns $E_{a}^{0}$ and $E_{a}$ in Table~1, provide a quantitative 
estimate of how much temperature effects may influence the calculation of migration energy
barriers in lithium FIC. In the present study, temperature effects account for as much as 
$35$-$80$\% of the final migration activation energies (straightforwardly estimated as $|E_{a} 
- E_{a}^{0}|/E_{a}$). Meanwhile, we have shown that, except for the Li$_{3}$OCl case, inclusion 
of temperature effects always brings into better agreement the calculated and measured migration 
energy barriers (Table~1). Therefore, we argue that AIMD simulations are strongly recommended
when pursuing accurate estimation of migration energy barriers (and of lithium diffusion mechanisms
as well), in spite of the much larger computational load associated to them as compared to zero-temperature
techniques. LiF, for example, illustrates very well the convenience of performing finite-temperature 
simulations. This material appears to be a very poor lithium-ion conductor \cite{yildirim15,pan15,soto18}, 
however, a relatively moderate migration energy barrier of $E_{a}^{0} = 0.68$~eV is calculated 
for it with $T = 0$~K methods. Such a value turns out to be quite similar to the $E_{a}^{0}$ obtained 
for other promising ionic conductors (e.g., $0.78$~eV for LiGaO$_{2}$ \cite{islam17}), hence
one could easily arrive at the wrong conclusion that LiF is a good superionic material. 

Interestingly, we appreciate that the calculated $E_{a}$'s are systematically higher than the 
corresponding values estimated at zero temperature. This observation could be interpreted 
such that thermally activated anion lattice vibrations tend to deplete lithium transport 
(since in zero-temperature $E_{a}^{0}$ calculations the anion sublattice vibrations are 
mostly neglected). However, other possible temperature effects associated to the diffusion
of mobile ions (e.g., entropic stabilization of diverse transition paths) and the complex 
interactions between cations and anions, cannot be disregarded. In the next subsection, we will 
comment in detail on the correlations between lithium conductivity ($E_{a}$ and $D_{0}$) 
and $\langle \omega \rangle$ that can be drawn from our AIMD results.

With regard to the estimation of average phonon frequencies, we note that the harmonic 
approximation may be not adequate for describing lithium FIC due to the high degree of 
anharmonicity that superionic phases normally present \cite{cazorla18b,chen15}. We 
have explicitly demonstrated this in Fig.\ref{fig1}, where the zero-temperature vibrational 
spectrum of non-stoichiometric Li$_{3}$N is shown to contain a considerable number of imaginary 
phonon frequencies, in contrast to the VDOS obtained under realistic $T \neq 0$~K 
conditions; we have checked that LiIO$_{3}$ and Li$_{3}$OCl behave in a very similar manner
(Supplementary Fig.7). In fact, the inherent limitations of the harmonic approximation 
may lead to misinterpretations of the vibrational stability of lithium FIC and to biased 
estimation of average phonon frequencies. For instance, we find that $\langle \omega \rangle$ 
is about $2$\% larger for Li$_{3}$N at zero temperature than at $T_{\rm room}$, and that such 
discrepancies propagate to the partial VDOS. It goes without saying that the presence of negative 
phonon frequencies in zero-temperature VDOS automatically invalidates any estimation of 
$\langle \omega \rangle$, due to violation of the fundamental assumptions from which harmonic 
approaches are deduced \cite{cazorla17c,cazorla13}. 
Correction of the explained computational artifacts (e.g., by using finite-temperature simulation
methods as done here) is very important for an improved interpretation of experiments, 
since partial VDOS normally cannot be resolved directly from measurements and consequently 
first-principles calculations are employed for that end \cite{muy18,krauskopf18}.

\subsection{Lattice dynamics versus migration activation energy and pre-exponential factor}
\label{susbsec:discussion1}
In Fig.\ref{fig8}a, we plot the $E_{a}$ and $\langle \omega \rangle$ results obtained for the 
superionic compounds considered in this study. A direct correlation between migration activation 
energies and average phonon frequencies cannot be established. For instance, Li$_{3}$N and LiGaO$_{2}$ 
present the smallest and largest $E_{a}$, respectively, with a large difference of $\sim 1$~eV, however 
the corresponding $\langle \omega \rangle$ turn out to be very similar. Likewise, one concludes 
the lack of any robust connection between either $\langle \omega \rangle_{\rm Li}$ or $\langle \omega 
\rangle_{\rm room}$ and $E_{a}$ (Figs.\ref{fig8}b,c and Table~1). Hence, the recently suggested 
correspondence between lattice softness and low activation energies in superionic argyrodites and 
Li$_{3}$PO$_{4}$-based LISICON \cite{kraft17,muy18} is not reproduced by our AIMD simulations and 
thereby should not be generalised to other families of lithium FIC. Actually, if LiGaO$_{2}$ was 
excluded from our analysis, we would arrive at the opposite conclusion that vibrationally rigid 
lattices, that is, larger $\langle \omega \rangle$, render smaller $E_{a}$ (see Fig.\ref{fig8}a). 

In previous sections we have shown that when lithium FIC are analysed individually, the large enhacements 
in ionic conductivity as induced by temperature are not accompanied by noticeable changes in VDOS. In particular, 
we have obtained large diffusivity variations of $\Delta D / D \sim 10^{1}-10^{3}$ and only minute vibrational 
changes of $\Delta \langle \omega \rangle / \langle \omega \rangle \sim 10^{-2}$. Hence, the generalised 
insensitivity of $\langle \omega \rangle$ for large $D$ fluctuations observed in all individual materials, 
already suggests the lack of a direct correlation between $E_{a}$ and $\langle \omega \rangle$ across 
different families of lithium FIC.  

Supplementary Fig.8 encloses the $D_{0}$ and $\langle \omega \rangle$ results obtained for the
superionic compounds investigated in this study. Once again, we cannot determine any rigorous
correspondence between the two represented quantities. We note that the calculated hopping attempt 
frequencies fluctuate between $0.01$ and $10$~THz whereas all the estimated $\langle \omega \rangle$ 
consistently amount to $\sim 10$~THz (Table~1). Therefore, lithium hoppings and lattice vibrations 
may operate at very different time scales, which could relate to the cause of their disconnection.

\subsection{Hopping attempt frequency versus $E_{a}$}
\label{subsec:prefactor}
The similarity of the $E_{a}$--$\langle \omega \rangle$ and $D_{0}$--$\langle \omega \rangle$ 
trends shown in Fig.\ref{fig8}a and Supplementary Fig.8 respectively, appears to suggest some sort 
of correlation between $E_{a}$ and $D_{0}$. In fact, the conventional hopping 
theory developed by Rice and Roth provides the well known, and usually reported, relationship 
$D_{0} \propto \sqrt{E_{a}}$ \cite{rice72}. However, we should note that conventional 
hopping theory was originally developed to understand ionic transport in AgI and other analogous 
type-I FIC \cite{hull04,cazorla17a}, in which the superionic transition is accompanied by a first-order 
structural transformation affecting the static sublattice (in contrast to lithium FIC, typically 
referred to as type-II, in which the superionic transition normally is of second-order type and 
mostly affects the mobile ions \cite{hull04,cazorla17a}). Moreover, a recent experimental study 
by Muy \emph{et al.} has provided solid evidence showing that the actual interplay between $D_{0}$ 
on $E_{a}$ may be more complex than previously thought \cite{muy18b}.

Figure~\ref{fig9}a reports the $D_{0}$ and $E_{a}$ results that we have obtained in this study. 
In this case, the two represented quantities appear to be correlated since larger migration 
activation energies systematically are accompanied by larger pre-exponential factors. However, we
note that the usually reported relationship $D_{0} \propto \sqrt{E_{a}}$ \cite{rice72} clearly does 
not pertain here (we recall that the $y$-axes in Fig.\ref{fig9} are in logarithmic scale). Rather, 
the dependence of $D_{0}$ on $E_{a}$ appears to be exponential like. Recently, Muy \emph{et al.} have 
shown that the diffusion pre-exponential factors of a large number of LISICON compounds seem
to follow the Meyer-Neldel rule $D_{0} \propto \exp{\left(E_{a}/\langle \omega \rangle\right)}$ 
\cite{muy18b}. In fact, when we represent the computed diffusion pre-exponential factors as a 
function of the quantity $E_{a}/\langle \omega \rangle$ we find a perfect agreement with the 
Meyer-Neldel rule within our numerical uncertainties (see linear fit in Fig.\ref{fig9}b). 
Our theoretical findings confirm Muy \emph{et al.}'s conclusions reported in work \cite{muy18b} 
and demonstrate that Rice and Roth's hopping theory in general is not adequate for describing 
lithium FIC (as it could have been foreseen, see explanations in previous paragraph).     

The results and discussions presented thus far let us to conclude the following: lattice softness 
can be identified with enhanced lithium diffusivity but only within families of superionic 
materials presenting very similar migration activation energies, due to superior $D_{0}$ (as 
given by the Meyer-Neldel rule). We should note that according to our zero-temperature and AIMD 
simulations lithium partial occupancy in FIC can be identified with larger anharmonicity, or 
equivalently, smaller $\langle \omega \rangle$. For instance, the average phonon frequency calculated 
for non-stoichiometric Li$_{3}$N at finite temperatures is about $1$\% lower than the estimated for 
the analogous stoichiometric system. Hence, lattice softness may indeed be a key factor for better 
understanding the ionic transport differences between chemically similar stoichiometric and 
non-stoichiometric lithium FIC \cite{muy18b}.

\section{Conclusions}
\label{sec:conclusions}
We have performed a comprehensive first-principles study of several lithium FIC at finite temperatures.
Based on our AIMD results, it has not been possible to establish any direct correlation between
either $E_{a}$ or $D_{0}$ and $\langle \omega \rangle$, in disagreement with recent experimental 
findings reported for superionic argyrodites and Li$_{3}$PO$_{4}$-based LISICON. Nevertheless, the 
three quantities of interest appear to be related by the Meyer-Neldel rule, in accordance with recent 
measurements; hence, on a general scale it is possible to identify lattice softness with enhanced 
ionic conductivity but only within families of FIC presenting very similar migration activation energies, 
owing to an increase in the hopping attempt frequency. Interestingly, we have shown that the spectra 
of lattice vibrations in lithium FIC generally are very insensitive to temperature changes, in 
contrast to what is observed for ionic transport. On the technical side, we have demonstrated 
that zero-temperature methods present some inherent limitations for describing Li-based FIC. In particular, 
migration activation energies can be seriously underestimated due to the neglection of temperature effects, 
and harmonic approaches may be ill-defined due to the prominent role of anharmonicity in FIC. We hope that 
our theoretical findings will help at establishing physically meaningful relationships between ionic 
transport and simple materials descriptors in lithium FIC. Also, we expect to promote a wider use 
of finite-temperature approaches in first-principles modeling of fast-ion conductors.

\section*{Acknowledgments}
This research was supported under the Australian Research Council's
Future Fellowship funding scheme (No. FT140100135). Computational resources
and technical assistance were provided by the Australian Government and the
Government of Western Australia through the National Computational Infrastructure
(NCI) and Magnus under the National Computational Merit Allocation Scheme and
The Pawsey Supercomputing Centre.

\end{document}